\documentclass[a4paper,10pt]{article}
\usepackage{amscd,amsfonts,amsmath}

\begin{document}
\title{The Inhomogeneous Invariance Quantum Supergroup of Supersymmetry Algebra}
\author{Azmi Ali Alt\i nta\d{s}\footnote{e-mail:ali.altintas@boun.edu.tr; fax:+902122872466}\;, Metin Ar\i k
\footnote{e-mail:metin.arik@boun.edu.tr}\\
{\small Physics Department, Bo\u{g}azi\c{c}i University, 34342
Bebek, Istanbul, Turkey}}
\date{\today}\maketitle
Pacs: 02.20.Uw, 11.30.Pb

\begin{abstract}
We consider an inhomogeneous quantum supergroup which leaves
invariant a supersymmetric particle algebra. The quantum
sub-supergroups of this inhomogeneous quantum supergroup are
investigated.
\end{abstract}

\section{Introduction}
Symmetry transformations based on Lie groups and Lie algebras are
most known in all areas of physics. Symmetry transformations are
algebraic objects and they contain Lie groups and Lie algebras as
special cases. Theory of quantum integrable systems has initiated
a new type of symmetry and mathematical objects called quantum
groups. The quantum groups are related to usual Lie groups as
quantum mechanics is related to its classical limit\cite{qg}.

Since the method of derivation of a quantum group from a group is
to make deformations on it, almost all examples of quantum groups
considered in physics have been deformations of ordinary groups
depending on one or more parameters. However, one can build up a
matrix whose elements satisfy Hopf algebra axioms just as the
quantum groups which are derived from ordinary groups.\cite{ali}
In this paper we use the term  ``invariance quantum supergroup" to
describe a Hopf superalgebra such that a supersymmetry algebra
forms a right module of the Hopf superalgebra.

An ordinary Lie algebra is defined over the field of complex numbers $\mathbb{C}$, however if an algebra is
defined over a vector space which has bosonic and fermionic elements, the algebra is called a
superalgebra\cite{susy}. An algebraic object which generalizes Lie superalgebras and their supergroups, is
called a quantum supergroup, more commonly Hopf superalgebra. Two algebras in a braided category have a natural
braided tensor product structure\cite{Majid}. Thus, in a superalgebra coproduct is defined via a braided tensor
product.

The Standard Model does not explain some aspects of cosmology,
concerning the large scale universe. For example, the Standard
Model can not explain the dark matter. However, supersymmetry
suggests explanations. The basic idea of supersymmetry is that the
equations which represent basic laws of nature do not change if
certain particles in the equations are interchanged with one
another\cite{susy2}. These equations have a supersymmetry because
they contain both fermionic and bosonic elements. These are called
SUSY algebras. The simplest SUSY algebra contains n bosons and n
fermions and these commute among each other. This algebra is
called the associative unital super algebra.

Since the associative unital superalgebra contains bosonic and
fermionic creation and annihilation operators one can build up a
noncommutative (NC) QFT. It is well known that QFT on 4- dimensional
NC space-time is invariant under the $SO(1,1)XSO(2)$ subgroup of the
Lorentz group.\cite{ch} However representation of this group is
different from the representation of Lorentz group. Using the notion
of twisted Poincar$\acute{e}$ symmetry one can show that the
interpretation of NC coordinates is extended from Lie algebra
framework to Hopf algebras. A consistent frame for NC QFT can be
realized in terms of twisted Poincar$\acute{e}$
symmetry.\cite{ch,ch2}

For an ordinary particle algebra, its quantum inhomogeneous
invariance quantum group has been discovered. For the fermion
algebra it is $FIO(2d)$ and for the boson algebra it is $BISp(2d)$
\cite{G,FIO,Bay} where d is number of fermions(bosons). In this
paper we look for an inhomogeneous quantum supergroup which leaves
the associative unital superalgebra invariant.

 \section{FBIOSp(2n$\mid$2m)}
  The associative unital superalgebra  contains fermions and bosons. Since the algebra does not change if the particles
interchanged with one to another, the associative unital superalgebra is an example of SUSY algebra. Using the
definition of graded commutator the commutation relations of the particle algebra can be written explicitly as
\begin{equation}\label{alg}
\begin{array}{cc}
\left\{f_{i},f_{j}\right\}=0 &
\left\{f_{i},f_{j}^{\ast}\right\}=\delta_{ij},\\
\left[b_{k},b_{l}\right]=0 & \left[b_{k},b_{l}^{\ast}\right]=\delta_{kl}\\
\left[f_{i},b_{k}\right]=0& \left[f_{i},b_{k}^{\ast}\right]=0,\\
\end{array}
\end{equation}
here $f_i$'s are fermion annihilation operators and $b_k$'s are
boson annihilation operators. In the algebra $i=1,\cdots, n$ and
$k=1,\cdots,m$ where $n$ is number of fermions, $m$ is number of
bosons. Also hermitian conjugates of the commutation relations are
valid. With respect to the commutation relations the associative
unital superalgebra is also a Lie superalgebra\cite{unital}. It is
well known that the supergroup $OSP(2n|2m)$ acting homogeneously
on the associative unital superalgebra leaves the commutation
relations (\ref{alg}) invariant\cite{rit2}. Here we would like to
consider an inhomogeneous transformation on bosons and fermions
and want the algebra to remain invariant under the transformation.
We write transformed fermionic and bosonic creation and
annihilation operators
\begin{equation}
\left(
\begin{array}{c}
f_i^{\prime } \\
f_i^{\ast ^{\prime }} \\
b_k^{\prime } \\
b_k^{\ast ^{\prime }} \\
1
\end{array}
\right) =\left(
\begin{array}{ccccc}
\alpha_{ij}& \beta_{ij}& a_{il}&c_{il}&\gamma_{i}\\
\beta_{ij}^{\ast}&\alpha_{ij}^{\ast}&c_{il}^{\ast}&a_{il}^{\ast}&\gamma_{i}^{\ast}\\
d_{kj}&e_{kj}&\varepsilon_{kl}&\theta_{kl}&\phi_{k}\\
\varepsilon_{kj}^{\ast}&\theta_{kj}^{\ast}&d_{kl}^{\ast}&e_{kl}^{\ast}&\phi_{k}^{\ast}\\
0&0&0&0&1
\end{array} \right) \dot{\otimes}\left(
\begin{array}{c}
f_j \\
f_j^{\ast}  \\
b_l\\
b_l^{\ast}\\ 1
\end{array}
\right).
\end{equation}
The transformation matrix T is given in terms of sub-matrices by
\begin{equation}T=\left(
\begin{array}{ccccc}
\alpha_{ij}& \beta_{ij}& a_{il}&c_{il}&\gamma_{i}\\
\beta_{ij}^{\ast}&\alpha_{ij}^{\ast}&c_{il}^{\ast}&a_{il}^{\ast}&\gamma_{i}^{\ast}\\
d_{kj}&e_{kj}&\varepsilon_{kl}&\theta_{kl}&\phi_{k}\\
\varepsilon_{kj}^{\ast}&\theta_{kj}^{\ast}&d_{kl}^{\ast}&e_{kl}^{\ast}&\phi_{k}^{\ast}\\
0&0&0&0&1
\end{array} \right)=\left(\begin{array}{ccc}
\alpha&A&\Gamma\\
D&\epsilon&\Phi\\
0&0&1
\end{array}
\right)=\left(
\begin{array}{c|c}
H & K \\\hline 0 & 1
\end{array}
\right),
\end{equation}
\begin{displaymath}
 \begin{array}{cc}
 \alpha=\left(
\begin{array}{cc}
\alpha_{ij}&\beta_{ij}\\
\beta_{ij}^{\ast}&\alpha_{ij}^{\ast}\end{array}\right),& A=\left(
\begin{array}{cc}
a_{il}&c_{il}\\
c_{il}^{\ast}&a_{il}^{\ast}\\
\end{array}\right),\\
 D=\left(\begin{array}{cc}
d_{kj}&e_{kj}\\
e_{kj}^{\ast}&d_{kj}^{\ast}\end{array}\right),&
\epsilon=\left(\begin{array}{cc}
\varepsilon_{kl}&\theta_{kl}\\
\theta_{kl}^{\ast}&\varepsilon_{kl}^{\ast}
\end{array}\right),\\
\Gamma=\left(\begin{array}{c}
\gamma_{i}\\
\gamma_{i}^{\ast}\end{array} \right), &
\Phi=\left(\begin{array}{c}
\phi_{k}\\
\phi_{k}^{\ast}
\end{array}\right).
\end{array}
\end{displaymath}
Where $H$ is the homogeneous and $K$ is the inhonogeneous part of
the matrix T. Here dimensions of the sub-matrices are:
$dim(\alpha)=(2n)\times (2n)$, $dim(A)=(n+m)\times(n+m)$,
$dim(D)=(m+n)\times (m+n)$ and $dim(\epsilon)=(2m)\times (2m)$.
$\Gamma$ is a $2n\times 1$ and $\Phi$ is a $2m\times 1$
dimensional matrix. First of all, we should mention that the
submatrices $A,\;D$ and $\Gamma$ have fermionic elements and the
others have bosonic elements. Secondly, we define a braided tensor
product. This is given by:
\begin{displaymath}
(P\otimes Q)(R\otimes S)= (-1)^{deg(Q)deg(R)}PR\otimes QS.
\end{displaymath}
\begin{displaymath}
deg(X)= \left\{
\begin{array}{cc}
1 & \mbox{fermionic}, \\
0 & \mbox{bosonic}.\\
\end{array}
\right.
\end{displaymath}
We should note that the above definition of braided tensor product
is true for only the $Z_2$ graded case namely, the algebra has
only bosonic and fermionic elements.
  We use transformed operators in equation (\ref{alg}) and want the algebra
to remain unchanged. Therefore, we get some relations involving
the elements of the transformation matrix T. We should mention
that braided tensor product was used to find the relations.
\begin{equation}
\begin{array}{cc}
\left[\alpha_{ij},\alpha_{kl}\right]=0 &
\left[\alpha_{ij},A_{kl}\right]=0,\\
\left[\alpha_{ij},D_{kl}\right]=0 & \left[\alpha_{ij},\epsilon_{kl}\right]=0,\\
\left[\alpha_{ij},\Gamma_{k}\right]=0& \left[\alpha_{ij},\Phi_{k}\right]=0,\\
\left\{A_{ij},D_{kl}\right\}=0&\left[A_{ij},\epsilon_{kl}\right]=0,\\
\left\{A_{ij},\Gamma_{k}\right\}=0&\left[A_{ij},\Phi_{k}\right]=0,\\
\left[D_{ij},\epsilon_{kl}\right]=0&\left[D_{ij},\Gamma_{k}\right]=0,\\
\left[D_{ij},\Phi_{k}\right]=0&\left\{\epsilon_{ij},\Gamma_{k}\right\}=0,\\
\left[\epsilon_{ij},\Phi_{k}\right]=0&\left\{A_{ij},A_{kl}\right\}=0,\\
\left\{D_{ij},D_{kl}\right\}=0\\
\end{array},
\end{equation}
also there are relations between the elements of matrices $\Gamma$
and $\Phi$. These can be written as;
\begin{equation}
  \begin{array}{c}
  \left\{\gamma_{i},\gamma_{j}\right\}=-\alpha_{ik}\beta_{jk}-\beta_{ik}\alpha_{jk}-c_{il}a_{jl}+a_{il}c_{jl},\\
   \left\{\gamma_{i}^{\ast},\gamma_{j}^{\ast}\right\}=\delta_{ij}-\alpha_{ik}\alpha_{jk}^{\ast}-\beta_{ik}\beta_{jk}^{\ast}-
c_{il}c_{jl}^{\ast}+a_{il}a_{jl}^{\ast},\\
   \left[\gamma_{i},\phi_{j}\right]=\alpha_{ik}e_{jk}+\beta_{ik}d_{jk}-c_{il}\varepsilon_{jl}+a_{il}\theta_{jl},\\
\left[\gamma_{i},\phi_{j}^{\ast}\right]=\alpha_{ik}d_{jk}^{\ast}+\beta_{ik}e_{jk}^{\ast}-c_{il}\theta_{jl}^{\ast}+a_{il}\varepsilon_{jl}^{\ast},\\
\left[\phi_{i},\phi_{j}\right]=d_{ik}e_{jk}+e_{ik}d_{jk}-\theta_{il}\varepsilon_{jl}+\theta_{il}\varepsilon_{jl},\\
\left[\phi_{i},\phi_{j}^{\ast}\right]=\delta_{ij}+d_{ik}d_{jk}^{\ast}+e_{ik}e_{jk}^{\ast}-\theta_{il}\theta_{jl}^{\ast}+
\varepsilon_{il}\varepsilon_{jl}^{\ast}.\\
  \end{array}
\end{equation}

 We look for a  Hopf superalgebra such that under this
transformation equation (\ref{alg}) remains invariant. In order to
do this, one should first check the coproduct. We want the
elements of the matrix T to belong to a  Hopf superalgebra
$\mathcal{H}$ where the coproduct is given by a matrix
multiplication
\begin{equation}\label{cop}
\Delta(T)=T\dot{\otimes}T.
\end{equation}
One can see that coproduct of commutation relations are satisfied
by defining braided tensor product. The counit is

\begin{equation}\label{cou}
\varepsilon (T)=\mathit{I}.
\end{equation}
Finally, the antipode should be found.
\begin{equation}\label{inv}
S(T)=T^{-1}
\end{equation}
\begin{equation}
T^{-1}=\left(
\begin{array}{cc}
H^{-1} & -H^{-1}K \\
0 & 1
\end{array}
\right) .
\end{equation}

Now we should find the inverse of homogeneous part $H$ of the
transformation matrix. Since the matrix H is a supermatrix, to
find its inverse the sub-matrices $\alpha$ and $\epsilon$ should
be invertible. We use the same technique which is used to find the
inverse of a supermatrix. The inverse matrix can be written as;
\begin{equation}H^{-1}=
\left(\begin{array}{cc}
\alpha'&A'\\
D'&\epsilon'\\
\end{array}
\right),
\end{equation} and the elements of $H^{-1}$ are:
\begin{eqnarray}
\alpha'&=&\left(\alpha-A\epsilon^{-1}D\right)^{-1},\\
\epsilon'&=&\left(\epsilon-D\alpha^{-1}A\right)^{-1},\\
A'&=&-\alpha^{-1}A\epsilon'\;,\\
D'&=&-\epsilon^{-1}D\alpha'\;.
\end{eqnarray}
Since the elements of $\alpha$ and $\epsilon$ are commutative,
$\alpha^{-1}$ and $\epsilon^{-1}$ can be found using the ordinary
matrix inverse rule.

The braided coproduct, the counit and the antipode of the
transformation matrix T as given by Eqs (\ref{cop}-\ref{inv}) have
been constructed. Thus, the transformation matrix T is an element
of a quantum supergroup. We may call this quantum supergroup
$FBIOSp(2n|2m)$, the Fermionic-Bosonic Inhomogeneous
Orthosymplectic quantum supergroup.

To find the quantum subgroups of $FBIOSp(2n|2m)$,  we should
impose additional constraints such as;
\begin{enumerate}
\item
$-\alpha_{ik}\beta_{jk}-\beta_{ik}\alpha_{jk}-c_{il}a_{jl}+a_{il}c_{jl}=0,\\
\delta_{ij}-\alpha_{ik}\alpha_{jk}^{\ast}-\beta_{ik}\beta_{jk}^{\ast}-
c_{il}c_{jl}^{\ast}+a_{il}a_{jl}^{\ast}=0,\\
\alpha_{ik}e_{jk}+\beta_{ik}d_{jk}-c_{il}\varepsilon_{jl}+a_{il}\theta_{jl}=0,\\
d_{ik}e_{jk}+e_{ik}d_{jk}-\theta_{il}\varepsilon_{jl}+\theta_{il}\varepsilon_{jl}=0,\\
\delta_{ij}+d_{ik}d_{jk}^{\ast}+e_{ik}e_{jk}^{\ast}-\theta_{il}\theta_{jl}^{\ast}+
\varepsilon_{il}\varepsilon_{jl}^{\ast}=0$.\\
\item $\Gamma=\Phi=0.$
\item $A=D=0.$
\item $\alpha=\epsilon=0.$
\end{enumerate}
Applying the above relations on $FBIOSp(2n|2m)$ one can get the
quantum subgroups and these are:
\[
\begin{CD}
{FBIOSp{(2n|2m})} @>{\bf(i)}>> {IOSp(2n|2m)} @>{\bf (ii)}>> {OSp(2n|2m)} \\
@V{\bf (iii)}VV @V{\bf (iii)}VV @V{\bf (iii)}VV \\
{FIO(2n)\times BISp(2m)} @>{\bf (i)}>> {GrIO(2n)\times ISp(2m)} @>{\bf (ii)}>> {O(2n)\times Sp(2m)} \\
@V{\bf (iv)}VV \\
{SA(n|m)}
\end{CD}
\]
Here $IOSp(2n|2m)$ is that the Inhomogeneous Orthosymplectic
supergroup where the elements of inhomogeneous part have graded
commutation relations among themselves. $OSp(2n|2m)$ is the
Orthosymplectic supergroup \cite{ortho}. $FIO(2d)$ is the
inhomogeneous invariance quantum group of the fermion algebra
\cite{G,FIO}. $BISp(2d)$ is the inhomogeneous invariance quantum
group of the boson algebra \cite{Bay}. $GrIO(2d)$ is the
grassmanian inhomogeneous orthogonal group. $ISp(2d)$ is the
Inhomogeneous Symplectic group \cite{ig}. $SA(n|m)$ is the
associative unital superalgebra whose elements satisfy equation
(\ref{alg}). We should mention that $FIO(2n)\times BISp(2m)$ has
quantum group structure. The others have group structure.
\section{Conclusion}
Supersymmetry claims that the equations of the laws of nature
should be invariant under transformations which change all basic
particles into each other. To describe laws of nature,
supersymmetric models are more convenient although there is yet no
experimental evidence. According to supersymmetry, every particle
has a partner particle which is a fermion for a boson and a boson
for a fermion. Supersymmetry algebra describes the symmetry
between bosons and fermions.

Graded Lie algebras and their graded Lie groups are used in
various branches of theoretical physics such as supersymmetric
field theory \cite{rit}. Both in graded Lie algebras and in
supersymmetry a grading factor is defined. For quantum groups, the
same grading can be defined, in which case they are called quantum
supergroups or braided quantum groups.

In this paper, we have shown that for the associative unital
superalgebra, an inhomogeneous quantum invariance supergroup can
be constructed and we have investigated its quantum
sub-supergroups.

The important point is that $FBIOSp(2n|2m)$ does not have a
deformation parameter. Other examples of quantum groups which do
not contain a deformation parameter are FIO(2d,R) and $U_G(d +
1)$\cite{ali2}.

The well known supergroups $OSp(2n|2m)$ and $O(2n)\times Sp(2m)$
are sub-supergroups of homogeneous part of $FBIOSp(2n|2m).$ In
this paper, we have shown that there is a general structure which
contains all the well known supergroups $IOSp(2n|2m)$,
$OSp(2n|2m)$ and $O(2n)\times Sp(2m)$. This general structure is
called as $FBIOSp(2n|2m)$ and it is a quantum supergroup.

As a further work, one can search for the inhomogeneous invariance
quantum group of a NC QFT and look at whether any relation between
this inhomogeneous quantum group and twisted Poincar$\acute{e}$
symmetry.

\end{document}